\documentclass[journal]{IEEEtran}
\usepackage[utf8]{inputenc}
\usepackage[english]{babel}
\usepackage{epsfig}
\usepackage{amsfonts}
\usepackage{ifpdf}
\usepackage{multirow}
\usepackage{amssymb}
\usepackage{amsthm}
\usepackage{mathtools}
\usepackage{mathrsfs}
\usepackage{textgreek}
\usepackage{grffile}
\usepackage{xcolor}
\usepackage{bm}

\usepackage{setspace}

\usepackage[usestackEOL]{stackengine}

  \stackMath

\ifCLASSINFOpdf
\else
\fi
\usepackage{amsmath}
\usepackage{algorithmic}
\usepackage{algorithm}
\usepackage{graphicx}

\newtheorem{definition}{Definition}

\newcommand{\norm}[1]{\left\lVert#1\right\rVert}

\ifCLASSOPTIONcompsoc
  \usepackage[caption=false,font=normalsize,labelfont=sf,textfont=sf]{subfig}
\else
  \usepackage[caption=false,font=footnotesize]{subfig}
\fi

\usepackage{stfloats}
\begin{document}
%\title{Massive MIMO Enabled Non-Coherent Modulation With One-bit ADCs}

%\title{Joint Positioning and Communication with Reconfigurable Intelligent Surfaces and Time-Resistant   }

\title{On the Limits of Single Anchor Localization: Near-Field vs. Far-Field}
\author{Don-Roberts~Emenonye,
        Harpreet~S.~Dhillon,
        and~R.~Michael~Buehrer% <-this % stops a space
\thanks{D.-R. Emenonye, H. S. Dhillon and R. M.  Buehrer are with Wireless@VT,  Bradley Department of Electrical and Computer Engineering, Virginia Tech,  Blacksburg,
VA, 24061, USA. Email: \{donroberts, hdhillon, rbuehrer\}@vt.edu. The support of the US National Science Foundation (Grants ECCS-2030215 and CNS-2107276) is gratefully acknowledged. 
}
}

\maketitle
\IEEEpeerreviewmaketitle
\begin{abstract}
%Location awareness is being extended from localizing agents to acquiring a picture of the entire scene. This picture involves acquiring the position of agents,  along with the position and orientation of anchors. Furthermore, it requires synchronization and building maps of the propagation environment through feature extraction and landmark location estimation. 
It is well known that a single anchor  can be used to determine the position and orientation of an agent communicating with it. However, it is not clear what information about the anchor or the agent is necessary to perform this localization, especially when the agent is in the near-field of the anchor. Hence, in this paper, to investigate the limits of localizing an agent with some uncertainty in the anchor location,  we consider a wireless link consisting of source and destination nodes.  More specifically, we present a Fisher information theoretical investigation of the possibility of estimating different combinations of the source and destination's position and orientation from the signal received at the destination. To present a comprehensive study, we perform this Fisher information theoretic investigation under both the near and far field propagation models. One of the key insights is that while the source or destination's $3$D orientation can be jointly estimated with the source or destination's $3$D position in the near-field propagation regime, only the source or destination's $2$D orientation can be jointly estimated with the source or destination's $2$D position in the far-field propagation regime. Also, a simulation of the FIM indicates that in the near-field, we can estimate the source's $3$D orientation angles with no beamforming, but in the far-field, we can not estimate the source's $2$D orientation angles when no beamforming is employed.

\end{abstract}
\section{Introduction}
%Futuristic applications of wireless technology, such as autonomous driving, require an extension of the concepts and precepts of location awareness to the more encompassing concept of situational awareness \cite{10086654,cole2020parameter}. Unlike location awareness, which has been investigated extensively, both with reconfigurable intelligent surfaces (RIS) \cite{emenonye2022fundamentals,emenonye2022ris,8264743,9729782,9508872} and under a single-anchor framework \cite{garcia2017direct,8240645,8356190,8515231,fascista2021downlink,8755880,li2019massive,guerra2018single}, situation awareness is a relatively novel idea. Situation awareness involves estimating the position of agents, along with the position and orientation of anchors. Furthermore, it involves synchronization and building maps of the propagation environment through feature extraction and landmark location estimation. 
Recently, due to the ubiquitous deployment of multi-antenna base stations, single-anchor localization has been proposed and studied with \cite{8240645,8356190,8515231} and without a reconfigurable intelligent surface (RIS) \cite{emenonye2022fundamentals,emenonye2022ris,9508872}. Localization is usually performed under the assumption that the anchor location (position and orientation) is perfectly known \cite{zekavat2011handbook}. However, in practical systems, this assumption might not hold. For example, in scenarios where unmanned aerial vehicles (UAV) act as anchors, there could be inherent uncertainty in the locations of the UAVs 
 \cite{banagar2022fundamentals,9206092}. Another  example involves localization using RISs. RISs are being considered to aid localization by acting as virtual anchors; however, their ubiquitous deployment means that their locations can change (e.g., when they are placed on movable objects), resulting in uncertainty in their locations. Lastly, in indoor localization systems, the locations of the indoor anchors can easily be disturbed after deployment. Hence, in this paper, to investigate localization with anchor uncertainty, we present a Fisher information view of estimating different combinations of a source and destination's position and orientation under the near and far field propagation regimes.
\subsection{Prior Art}
Prior literature on single-anchor localization involves deriving the fundamental limits for the accuracy achievable in estimating the position and orientation of an agent \cite{8240645}. These bounds are extended to the case of $3$D localization of an agent in \cite{8356190}. In \cite{8515231}, the amount of information in the non-line of sight (NLOS) paths and their usefulness for localization is analyzed. The bounds of single-anchor localization with a RIS have been studied in \cite{emenonye2022fundamentals}. These bounds are extended to account for near-field propagation in \cite{emenonye2022ris,9508872}. In the context of anchor state uncertainty, localization has been investigated with and without a  RIS. In \cite{6684554}, the positioning problem in the presence of anchor uncertainty is studied, the resulting non-convex optimization problem is relaxed to a second-order cone programming problem, and semidefinite programming is applied. The authors in \cite{8194834} derive the geometric dilution of precision in the presence of anchor position uncertainty, and a trade-off is made between range errors and
position errors by applying the modified spring mass method. The anchor position offset and the agent's position are estimated in \cite{6746771} using the signal strength of the received signals. In \cite{7055921}, a rigorous investigation of the impact of anchor uncertainty on received signal strength-based localization techniques is presented. The bounds given in \cite{7055921} serve as lower bounds to the algorithm in \cite{6746771}. %Similar to \cite{7055921}, fundamental lower bounds are derived in \cite{shi2020sequential}, but for the case of localizing moving agents in the presence of anchor uncertainty. 
In \cite{9044729}, multipath propagation between the uncertain anchor and the agent is investigated, the error model of the anchor uncertainty is assumed, and importance sampling is used to obtain the agent's position. %In \cite{7784698}, cooperative localization is used to calibrate the uncertainty in the position of the anchors, and the framework is extended to include multiple anchors and agents in \cite{10086654}. 
Uncertainties are considered in the case of RIS-assisted localization in \cite{emenonye2022fundamentals,emenonye2022ris}. While the prior art primarily includes robust algorithms to handle uncertainty in anchors' position, a comprehensive Fisher information-based analysis on the estimation of the anchor orientation has yet to be studied. It is important to note the anchor orientation is particularly important as the localization of agents is now being considered with large antenna single anchors. Moreover, the effect of anchor location uncertainty has not been investigated under the near-field propagation regime.  
\subsection{Contributions}
In this paper, through the Fisher information matrix (FIM), we present a theoretical investigation of the limits of single-anchor localization by determining the combinations of positions and orientations of the source and destination nodes that can be estimated in the near and far field propagation regimes. Further, using the FIM, we present a lower bound for the source orientation and destination position accuracy. One key result from the FIM-based analysis is that in the near-field, the source or destination's $3$D orientations can be estimated jointly with either the source or destination's $3$D positions. Also, in the far-field, the source or destination's $2$D orientations can be estimated jointly with either the source or destination's $2$D positions. Another result is that while the presence of a beamforming matrix is not required in the near-field to  estimate the source's $3$D orientation angles,  a beamforming matrix is required in the far-field to estimate the source's $2$D orientation angles.

\textit{Notation:} 
the transpose operator is $(\cdot)^{\mathrm{T}}$; the hermitian transpose operator is $(\cdot)^{\mathrm{H}}$; the  submatrix in the matrix $\bm{V}$, with rows in the range, $g_1:v_1$,  and the columns  in the range $g_2:v_2$ is extracted using the operation $[\bm{V}]_{[g_1:v_1,g_2:v_2 ]}$;  $\operatorname{Tr}(\cdot)$ is the matrix trace operator; 
$\norm{\cdot}$ denotes the Euclidean norm  ; the positive definiteness of a matrix is characterized by $ \succ$  ;
the first derivative operator is $\nabla$ ; 
 the expectation operator with respect to the random vector $\bm{v}$ is $\mathbb{E}_{\bm{v}}\{\cdot\}$.
\section{System Model}
We consider a source with its centroid located at $\bm{p}_{B} = [x_B, y_B, z_B]^{\mathrm{T}}$, and its $b^{\text{th}}$ antenna element located at $\bm{s}_{b} = [x_b, y_b, z_b]^{\mathrm{T}}$. The location of the centroid is defined with respect to the global origin, while the location specified by $\bm{s}_{b}$  is defined with respect to $\bm{p}_{B}$.  

\begin{figure}[htb!]
\centering
%\subfloat[]{\includegraphics[height=1.8in, width= 1.5in]{Results/ber_ring_ratio_parallel.eps}
{\includegraphics[scale=0.3]{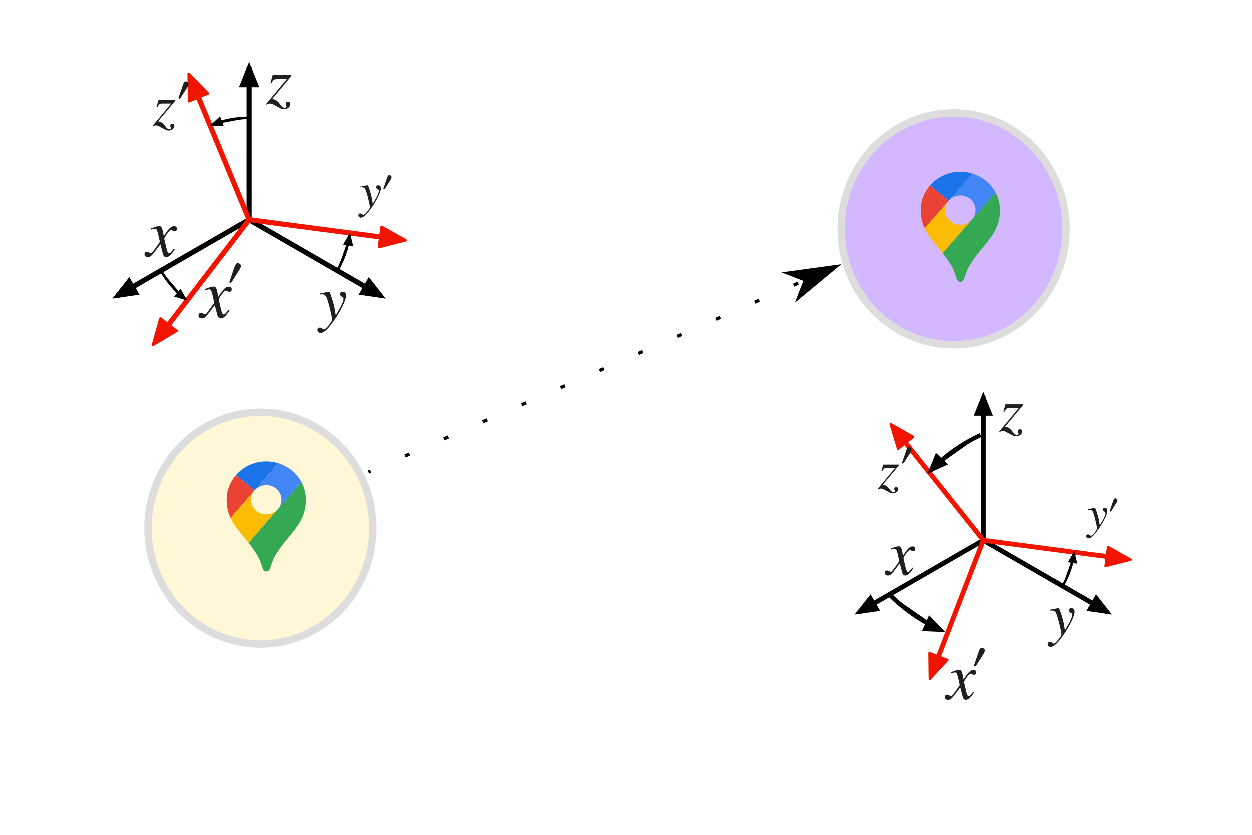}}
\caption{An illustration showing a source communicating with a destination.} 
\label{fig:Figure/system_model_final}
\end{figure}

This point $\bm{s}_b$ can also be written as $\bm{s}_b = \bm{Q}_B \Tilde{\bm{s}}_b$, where $\Tilde{\bm{s}}_{b} = [\tilde{x}_b, \tilde{y}_b, \tilde{z}_b]^{\mathrm{T}}$ is the previously known position of the antenna coordinate with respect to $\bm{p}_B$ before an orientation offset, $\bm{\Phi}_B =  [\alpha_{B}, \psi_{B}, \varphi_{B}]^{\mathrm{T}}$. The subsequent $3$D orientation matrix is defined as  $\bm{Q}_{B}$ \cite{lavalle2006planning}. There are $N_B$ antennas at the source, and each antenna can be described with respect to the global origin as $\bm{p}_b = \bm{p}_B +  \bm{s}_b$.  The destination is located at $\bm{p}_{U} = [x_U, y_U, z_U]^{\mathrm{T}}$, and its $u^{\text{th}}$ antenna element is located at $\bm{s}_{u} = [x_u, y_u, z_u]^{\mathrm{T}}$. The corresponding vectors, $\bm{p}_U$,  $\bm{s}_u$,  $\Tilde{\bm{s}}_u$ and $\bm{p}_u$ have similar definitions as the corresponding source's  vectors. Note that the orientation angles and the matrix related to the destination are denoted by 
 $\bm{\Phi}_{U}$ and $\bm{Q}_{U}$, respectively. %However, we assume that the  $\bm{s}_u  = \Tilde{\bm{s}}_u$, hence $\bm{Q}_{U} = \bm{I}$. 
The position of the destination's centroid located at ${\bm{p}}_{{U}}$ can be described in relation to the position of the  source's centroid located at ${\bm{p}}_{{B}}$ as ${\bm{p}}_{{U}} = {\bm{p}}_{{B}}     +d_{{B} {U}} \bm{\Delta}_{{B} {U}},$
%\begin{equation}
%%   \begin{aligned}
%{\bm{p}}_{{U}} &= {\bm{p}}_{{B}}     +d_{{B} {U}} \bm{\Delta}_{{B} {U}},
  %  \end{aligned}
%\end{equation}
where  $d_{{B} {U}}$ is the distance from point ${\bm{p}}_{{B}}$ to point ${\bm{p}}_{{U}}$ and $\bm{\Delta}_{{B} {U}}$ is the corresponding unit direction vector $\bm{\Delta}_{{B} {U}} = [\cos \phi_{{B} {U}} \sin \theta_{{B} {U}}, \sin \phi_{{B} {U}} \sin \theta_{{B} {U}}, \cos \theta_{{B} {U}}]^{\mathrm{T}}$. All points defined locally that describe the location of elements on the source antenna array with respect to the source's centroid can be written in the matrix form as
 $ {\bm{S}}_{B} = [{\bm{s}}_1, {\bm{s}}_2, \cdots, {\bm{s}}_{N_B} ].$ Similarly, the points defined locally that describe the location of elements on the destination antenna array with respect to the destination's centroid can be written in the matrix form as
 $ {\bm{S}}_{U} = [{\bm{s}}_1, {\bm{s}}_2, \cdots, {\bm{s}}_{N_U} ].$ Matrices $\Tilde{\bm{S}}_{B}$ and $\Tilde{\bm{S}}_{U}$ can be described similarly, by collecting  the appropriate vectors $\Tilde{\bm{s}}_b$ and $\Tilde{\bm{s}}_{u}$.
 \subsection{Signal Model}
 The communication from the source to the destination is achieved through the transmission of $T$ symbols from the source with $N_B$ transmit antennas to the destination with $N_U$ receive antennas. During each transmission, the source precodes a data stream, $\bm{x} \in \mathcal{C}^{N_D \times 1}$,  to the $N_B$ transmit antennas with a beamforming matrix $\bm{F}_{t} \in \mathcal{C}^{N_B \times N_D}$ under the constraint $\mathbb{E}\left\{\norm{\bm{x}}^2\right\}=1$. The signal received during the $t^{\text{th}}$ transmission is
\begin{equation}
\label{equ:receive_processing}
\begin{aligned}
\bm{y}_{t} =  \bm{H}_{}^{} \bm{F}_t \bm{x}
+ \bm{n}_{t}, =   \bm{\mu}^{}_{t}+  \bm{n}_{t}.
\end{aligned}
\end{equation}
In the above equation,  $\bm{\mu}^{}_{t}$  is the noise-free part (useful part) of the signal, and $\bm{n}_{t}\sim \mathcal{C}\mathcal{N}(0, N_0)$  represents the thermal noise local to the destination's antenna array. The element in the $u^{\text{th}}$ row and $b^{\text{th}}$ column of the channel matrix $\bm{H}$ is $[\bm{H}]_{[u,b]} = \beta e^{-j 2 \pi f_{c} \tau_{ bu}}$. Here, $\beta  = \beta_{\mathrm{R}} + j\beta_{\mathrm{I}}$ is the complex path gain, $f_c$ is the operating frequency, and $\tau_{ bu}$ is the propagation delay from the  $b^{\text{th}}$ transmit antenna located at $\bm{p}_b$ on the source's antenna array to the receive antenna located at $\bm{p}_u$ on the destination's antenna array. Now, the signal received at the destination's $u^{\text{th}}$ receive antenna during the $t^{\text{th}}$ transmission is
\begin{equation}
\label{equ:receive_processing_1}
\begin{aligned}
\bm{y}_{t,u} &=  \sum_{b = 1}^{N_B} \sum_{d = 1}^{N_D}[\bm{F}_t]_{[b,d]}^{} [\bm{x}]_{[d]}[\bm{H}]_{[u,b]}^{}  
+ \bm{n}_{t}.
\end{aligned}
\end{equation}
The definition of the delay given as $\tau_{ bu} = \frac{\norm{\bm{p}_{u} - \bm{p}_{b}}}{c}$ incorporates any potential spherical curvature wavefront present in the signal received at the destination. When the destination experiences substantial wavefront curvature, it is said to be located within the near-field propagation regime. It is important to note that at sufficiently larger distances  between the destination and the source, the spherical wavefront can be approximated by a plane wave. With this plane wave approximation, the delay can be approximated as $\tau_{bu} = \tau_{BU} + \Delta_{BU}^{\mathrm{T}} ({\bm{s}}_{u}- {\bm{s}}_{b}) / c.$ When this approximation holds, the destination is said to be located within the far-field propagation regime.  The boundary that defines the near and far field propagation regime is called the Fraunhofer distance. This Fraunhofer distance can be computed as $d_{\mathrm{f}}=2 D^2 / \lambda$     with $\lambda$ indicating the wavelength of the signal and  $D$  the maximum diameter among the source and destination surface diameters \cite{emenonye2022ris}. While, (\ref{equ:receive_processing}) and (\ref{equ:receive_processing_1}) adequately represent the signals received in the near-field, an approximation of signals received in the far-field can be written as
\begin{equation}
\label{proposition_equ:far_field_1}
\begin{aligned}
&\bm{y}_{t} =  \beta^{} \bm{a}_{UB}(\Delta_{BU}) \bm{a}_{BU}^{\mathrm{H}}(\Delta_{BU})  e^{-j 2 \pi f_{c}
\tau_{BU}}\bm{F}_t \bm{x} + \bm{n}_{t},
\end{aligned}
\end{equation}
where $\bm{a}_{BU}(\Delta_{BU}) = e^{-j 2 \frac{\pi}{\lambda} {\bm{S}}_{B}^{\mathrm{T}} \Delta_{BU} } $ and $\bm{a}_{UB}(\Delta_{BU}) = e^{-j 2 \frac{\pi}{\lambda} {\bm{S}}_{U}^{\mathrm{T}} \Delta_{BU} }$.
\subsection{Source and Destination Position  and Orientation Estimation}
In this letter, we provide the different combinations of source and destination position and orientation that can be estimated through the signals received across the $N_U$ antennas during the $T$ transmissions. We determine this by evaluating the FIM under the following parameterizations:   case I) $\bm{\eta} = [\bm{p}_{U}, \bm{\Phi}_{U}, \bm{\beta}
]^{\mathrm{T}}$, case II) $\bm{\eta} = [\bm{p}_{U}, \bm{\Phi}_{B}, \bm{\beta}
]^{\mathrm{T}}$, case III)  $\bm{\eta} = [\bm{p}_{B}, \bm{\Phi}_{U}, \bm{\beta}
]^{\mathrm{T}}$, and  case IV) $\bm{\eta} = [\bm{p}_{B}, \bm{\Phi}_{B}, \bm{\beta}
]^{\mathrm{T}}$. Here, $\bm{\beta} = [\beta_{\mathrm{R}}, \beta_{\mathrm{I}}]^{\mathrm{T}}$.  Note that the location parameters for each individual case can be collected into the vector $\bm{\zeta}$.  The FIM computations are carried under three scenarios: i) far-field model with beamforming, ii)  near-field model with no beamforming, and iii) near-field model with beamforming. Note that the case for using the far-field model with identity beamforming matrices across the $T$ transmissions is not possible.  This is because the joint estimation of the source orientation, $\bm{\Phi}_{B}$, and $\bm{\beta}$ is not feasible under this condition (see Appendix \ref{appendix_joint_estimation}).
\section{Information in the Received signal}
To analyze the amount of location information present in the received signal, we introduce the mathematical definition of the FIM for an unknown parameter vector, $\bm{\eta}$, in the following definition.

\begin{definition}
\label{definition_FIM_1_fundamentals}
Based on a set of observations $\mathbf{y}$, the Fisher information of a parameter vector, $\bm{\eta}_{}$,  is written as
\begin{equation}
\label{equ:definition_FIM_1}
\begin{aligned}
\mathbf{J}_{ \bm{\eta}_{}} &\triangleq 
-\mathbb{E}_{\mathbf{y} {}}\left[\frac{\partial^{2} \ln \chi(\mathbf{y}_{}|  \bm{\eta}_{} )}{\partial \bm{\eta}_{} \partial \bm{\eta}_{}^{\mathrm{T}}}\right] 
\end{aligned}
\end{equation}
where $\mathbb{E}_{\nu}$ is expectation taken over the random variable $\nu$, $\chi(\mathbf{y}_{}|  \bm{\eta}_{} )$ is the likelihood of $\bm{y}$ conditioned on $\bm{\eta}$. We note that the error covariance matrix of an unbiased estimate, $\hat{\bm{\eta}}$, of an unknown parameter vector, ${\bm{\eta}}$ satisfies the following information inequality $
\mathbb{E}_{\bm{y}}\left\{(\hat{\boldsymbol{\eta}}-\boldsymbol{\eta})(\hat{\boldsymbol{\eta}}-\boldsymbol{\eta})^{\mathrm{T}}\right\} \succeq \mathbf{J}_{  \bm{\eta}}^{-1}.
$
\end{definition}

The FIM for the parameter vector $\bm{\eta} = [\bm{p}_{U}, \bm{\Phi}_{B}, \bm{\beta}
]^{\mathrm{T}}$ has the following structure
\begin{equation}
\label{equ:FIM_parameter_matrix}
\begin{aligned}
\mathbf{J}_{ \bm{\eta}_{}}
 \triangleq\left[\begin{array}{cccc}
\mathbf{J}_{\bm{p}_{U}\bm{p}_{U}} & \mathbf{J}_{\bm{p}_{U}\bm{\Phi}_{B}} & \mathbf{J}_{\bm{p}_{U}{\bm{\beta}_{\mathrm{R}}}} & \mathbf{J}_{\bm{p}_{U}{\bm{\beta}_{\mathrm{I}}}} \\
\mathbf{J}_{\bm{\Phi}_{B}\bm{p}_{U}} & \mathbf{J}_{\bm{\Phi}_{B}\bm{\Phi}_{B}} & \mathbf{J}_{\bm{\Phi}_{B}{\bm{\beta}_{\mathrm{R}}}} & \mathbf{J}_{\bm{\Phi}_{B}{\bm{\beta}_{\mathrm{I}}}} \\
\mathbf{J}_{{\bm{\beta}_{\mathrm{R}}}\bm{p}_{U}} & \mathbf{J}_{{\bm{\beta}_{\mathrm{R}}}\bm{\Phi}_{B}} & \mathbf{J}_{{\bm{\beta}_{\mathrm{R}}}{\bm{\beta}_{\mathrm{R}}}} & \mathbf{J}_{{\bm{\beta}_{\mathrm{R}}}{\bm{\beta}_{\mathrm{I}}}} \\
\mathbf{J}_{{\bm{\beta}_{\mathrm{I}}}\bm{p}_{U}} & \mathbf{J}_{{\bm{\beta}_{\mathrm{I}}}\bm{\Phi}_{B}} & \mathbf{J}_{{\bm{\beta}_{\mathrm{I}}}{\bm{\beta}_{\mathrm{R}}}} & \mathbf{J}_{{\bm{\beta}_{\mathrm{I}}}{\bm{\beta}_{\mathrm{I}}}} 
\end{array}\right] \in \mathcal{R}^{8 \times 8}.
\end{aligned}
\end{equation}
The submatrices in the above matrix can be computed using $\mathbf{J}_{\bm{\eta}_{{\mathbf{v}_1}} \bm{\eta}_{{\mathbf{v}_2}}} \triangleq \frac{2}{\sigma^2} \sum_{t=1}^{{T}} \Re\left\{\frac{\partial \bm{{\mu}_{t}}^{\mathrm{H}}}{\partial {\bm{\eta}_{}}_{\mathbf{v}_1 }} \frac{\partial \bm{{\mu}{t}}^{{}}}{\partial {\bm{\eta}_{}}_{\mathbf{v}_2 }}\right\}$   where $\bm{\eta}_{\mathbf{v}_1} \in \bm{\eta}  $, $\bm{\eta}_{\mathbf{v}_2} \in \bm{\eta}  $ are both dummy variables, and $1/\sigma^2$ is the SNR which incorporates the pathloss and composite noise power. The required first derivatives are presented in the following sections.
\subsection{First Derivatives under the Far-Field Model}
The first derivative of the useful part of the received signal with respect to $\nu \in [\bm{p}_{B}, \bm{p}_{U}]$ under the far-field model is
$$
 \nabla_{\nu} {\bm{\mu}_{t,u}}  =\beta^{} e^{-j 2 \frac{\pi}{\lambda} {\bm{s}}_{u}^{\mathrm{T}} \Delta_{BU} } \bm{a}_{BU}^{\mathrm{H}}(\Delta_{BU}) \bm{K}_{\nu} e^{-j 2 \pi f_{c}
\tau_{BU}}\bm{F}_t \bm{x},
$$
where $\bm{K}_{\nu}$ is expressed in (\ref{equ:k_nu}).
\begin{figure*}
\begin{align}
\label{equ:k_nu}
\bm{K}_{\nu} & = \text{diag}\Bigg[-\frac{j2\pi}{\lambda} \bigg(\bm{s}_{u}^{\mathrm{T}} \nabla_{\nu} {\Bigg[\frac{\bm{p}_{U} - \bm{p}_{B}}{d_{BU}}\Bigg]} - \nabla_{\nu}{\Bigg[\frac{\bm{p}_{U} - \bm{p}_{B}}{d_{BU}}\Bigg]}^{\mathrm{T}} \bm{S}_{B}    +   \nabla_{\nu} d_{BU} \bigg)\Bigg], \\
\nabla_{{\bm{p}^{}_{B}} }d_{BU} & = -1 \times {\Bigg[\frac{\bm{p}_{U} - \bm{p}_{B}}{d_{BU}}\Bigg]}, \; \;
\nabla_{{\bm{p}^{}_{B}} } {\Bigg[\frac{\bm{p}_{U} - \bm{p}_{B}}{d_{BU}}\Bigg]}   = \frac{-d_{BU} - (\bm{p}_{U} - \bm{p}_{B})\nabla_{{\bm{p}^{}_{U}} }d_{BU}}{d_{BU}^2}, \\
\nabla_{{\bm{p}^{}_{U}} }d_{BU} & = {\Bigg[\frac{\bm{p}_{U} - \bm{p}_{B}}{d_{BU}}\Bigg]}, \; \;
\nabla_{{\bm{p}^{}_{U}} } {\Bigg[\frac{\bm{p}_{U} - \bm{p}_{B}}{d_{BU}}\Bigg]}   = \frac{d_{BU} - (\bm{p}_{U} - \bm{p}_{B})\nabla_{{\bm{p}^{}_{U}} }d_{BU}}{d_{BU}^2}.
\end{align}
\end{figure*}
The first derivatives of the useful part of the received signal with respect to $\nu \in \bm{\Phi}_{B}$ and $\nu \in \bm{\Phi}_{U}$ under the far-field model are
$$
 \nabla_{\nu} {\bm{\mu}_{t}}  =\beta^{} \bm{\tilde{P}}_{\nu} \bm{a}_{UB}(\Delta_{BU}) \bm{a}_{BU}^{\mathrm{H}}(\Delta_{BU})  e^{-j 2 \pi f_{c}
\tau_{BU}}\bm{F}_t \bm{x},
$$
$$
 \nabla_{\nu} {\bm{\mu}_{t,u}}  =\beta^{} e^{-j 2 \frac{\pi}{\lambda} {\bm{s}}_{u}^{\mathrm{T}} \Delta_{BU} } \bm{a}_{BU}^{\mathrm{H}}(\Delta_{BU}) \bm{P}_{\nu} e^{-j 2 \pi f_{c}
\tau_{BU}}\bm{F}_t \bm{x},
$$
respectively, where
$$
\begin{aligned}
   \tilde{\bm{P}}_{\nu} & =  \text{diag}\Bigg[-\frac{j2\pi}{\lambda}(\nabla_{\nu}\bm{S}_{u})^{\mathrm{T}}  {\Bigg[\frac{\bm{p}_{U} - \bm{p}_{B}}{d_{BU}}\Bigg]} \Bigg], \\
    \bm{P}_{\nu} & = \text{diag}\Bigg[\frac{j2\pi}{\lambda} {\Bigg[\frac{\bm{p}_{U} - \bm{p}_{B}}{d_{BU}}\Bigg]}^{\mathrm{T}} \nabla_{\nu}\bm{S}_{B}   \Bigg].
\end{aligned}
$$
Also, $\nabla_{{\bm{\Phi}^{}_{B}} } \bm{S}_{B}   = \nabla_{{\bm{\Phi}^{}_{B}} } \bm{Q}_{B}\Tilde{\bm{S}}_{B}$ and $
\nabla_{{\bm{\Phi}^{}_{U}} } \bm{S}_{U}  = \nabla_{{\bm{\Phi}^{}_{U}} } \bm{Q}_{U}\Tilde{\bm{S}}_{U}. $
Finally, the first derivative of the useful part of the received signal with respect to complex path gain under the far-field model is
$$
\begin{aligned}
\nabla_{{\bm{\beta}_{\mathrm{R}}} } \bm{\mu}_{t} & = \bm{a}_{UB}(\Delta_{BU}) \bm{a}_{BU}^{\mathrm{H}}(\Delta_{BU}) \bm{F}_{t}\bm{x}  e^{-j 2 \pi f_{c}
\tau_{BU}}, \\
\nabla_{{\bm{\beta}_{\mathrm{I}}} } \bm{\mu}_{t} & = j\bm{a}_{UB}(\Delta_{BU}) \bm{a}_{BU}^{\mathrm{H}}(\Delta_{BU}) \bm{F}_{t}\bm{x}  e^{-j 2 \pi f_{c}
\tau_{BU}}. \\
\end{aligned}
$$
The above first derivatives are used to compute the submatrices with a similar structure as that shown in 
(\ref{equ:FIM_parameter_matrix}) when the far-field model is used.
\subsection{First Derivatives under the Near-Field Model}
The first derivatives of the useful part of the received signal with respect to $\bm{\eta}$ under the near-field model are
$$
%\frac{\partial \bm{{\mu}_{t}}^{\mathrm{H}}}{\partial {\bm{p}_{U}}}
\begin{aligned}
%\nabla_{{\bm{p}_{B}} } {\bm{\mu}_{t,u}} & = (-j2\pi f_c) \beta \sum_{b = 1}^{N_B} \nabla_{{\bm{p}_{B}} } {\tau_{bu}}  \sum_{d = 1}^{N_D}[\bm{F}_t]_{[b,d]}^{} [\bm{x}]_{[d]}e^{-j 2 \pi f_{c}
%\tau_{bu}} 
 %, \\
\nabla_{{\bm{p}_{U}} } {\bm{\mu}_{t,u}} & = (-j2\pi f_c) \beta \sum_{b = 1}^{N_B} \nabla_{{\bm{p}_{U}} } {\tau_{bu}}  \sum_{d = 1}^{N_D}[\bm{F}_t]_{[b,d]}^{} [\bm{x}]_{[d]}e^{-j 2 \pi f_{c}
\tau_{bu}}, \\ 
\nabla_{{\bm{p}_{B}} } {\bm{\mu}_{t,u}} & = (-j2\pi f_c) \beta \sum_{b = 1}^{N_B} \nabla_{{\bm{p}_{B}} } {\tau_{bu}}  \sum_{d = 1}^{N_D}[\bm{F}_t]_{[b,d]}^{} [\bm{x}]_{[d]}e^{-j 2 \pi f_{c}
\tau_{bu}}, \\ 
\nabla_{{\bm{\Phi}_{U}} } {\bm{\mu}_{t,u}} & = (-j2\pi f_c) \beta \sum_{b = 1}^{N_B} \nabla_{{\bm{\Phi}_{U}} } {\tau_{bu}}  \sum_{d = 1}^{N_D}[\bm{F}_t]_{[b,d]}^{} [\bm{x}]_{[d]}e^{-j 2 \pi f_{c}
\tau_{bu}}, \\
\nabla_{{\bm{\Phi}_{B}} } {\bm{\mu}_{t,u}} & = (-j2\pi f_c) \beta \sum_{b = 1}^{N_B} \nabla_{{\bm{\Phi}_{B}} } {\tau_{bu}}  \sum_{d = 1}^{N_D}[\bm{F}_t]_{[b,d]}^{} [\bm{x}]_{[d]}e^{-j 2 \pi f_{c}
\tau_{bu}} .
\end{aligned}
$$
%The derivatives related to ${\bm{p}_{B}}$ and ${\bm{\Phi}_{U}} $ can be  similarly obtained.
% \nabla_{{\bm{\Phi}_{U}} } {\bm{\mu}_{t,u}} & = (-j2\pi f_c) \beta \sum_{b = 1}^{N_B} \nabla_{{\bm{\Phi}_{U}} } {\tau_{bu}}  \sum_{d = 1}^{N_D}[\bm{F}_t]_{[b,d]}^{} [\bm{x}]_{[d]}e^{-j 2 \pi f_{c}
%\tau_{bu}} 
% , \\
 $$
%\frac{\partial \bm{{\mu}_{t}}^{\mathrm{H}}}{\partial {\bm{p}_{U}}}
\begin{aligned}
\nabla_{{\bm{\beta}_{\mathrm{R}}} } \bm{\mu}_{t,u} & =  \sum_{b = 1}^{N_B} \sum_{d = 1}^{N_D}[\bm{F}_t]_{[b,d]}^{} [\bm{x}]_{[d]}e^{-j 2 \pi f_{c}
\tau_{bu}} 
 , \\
\nabla_{{\bm{\beta}_{\mathrm{I}}} } \bm{\mu}_{t,u} & = j \sum_{b = 1}^{N_B}   \sum_{d = 1}^{N_D}[\bm{F}_t]_{[b,d]}^{} [\bm{x}]_{[d]}e^{-j 2 \pi f_{c}
\tau_{bu}}.
\end{aligned}
$$
 Here, $\nabla_{{\bm{p}_{B}} } {\tau_{bu}}  =  \nabla_{{\bm{p}_{B}} } {d_{bu}} / c $, $\nabla_{{\bm{p}_{U}} } {\tau_{bu}}  =  \nabla_{{\bm{p}_{U}} } {d_{bu}} / c $,
 $\nabla_{{\bm{\Phi}_{U}} } {\tau_{bu}}  =  \nabla_{{\bm{\Phi}_{U}} } {d_{bu}} / c$, 
 $ \nabla_{{\bm{\Phi}_{B}} } {\tau_{bu}}  =  \nabla_{{\bm{\Phi}_{B}} } {d_{bu}} / c$,
 $\nabla_{{\bm{p}_{B}} } {d_{bu}}  =  -\frac{\bm{p}_{u} - \bm{p}_{b}}{d_{bu}}$,
 $\nabla_{{\bm{p}_{U}} } {d_{bu}}  =  \frac{\bm{p}_{u} - \bm{p}_{b}}{d_{bu}}$, $\nabla_{{\bm{\Phi}_{B}} } {d_{bu}} 
   = - \frac{ (\bm{p}_{u} - \bm{p}_{b})^{\mathrm{T}}}{d_{bu}}(\nabla_{{\bm{\Phi}^{}_{B}} } \bm{Q}_{B}\Tilde{\bm{s}}_{b}),$ and $\nabla_{{\bm{\Phi}_{U}} } {d_{bu}} 
   =  \frac{ (\bm{p}_{u} - \bm{p}_{b})^{\mathrm{T}}}{d_{bu}}(\nabla_{{\bm{\Phi}^{}_{U}} } \bm{Q}_{U}\Tilde{\bm{s}}_{u}).$
The above first derivatives are used to compute the submatrices with a similar structure as that shown in
(\ref{equ:FIM_parameter_matrix}) when the near-field model is used.
 After computing $\mathbf{J}_{\bm{\eta}}$, to focus on the available information concerning the location parameters, we present a mathematical description of the EFIM. 
 \begin{definition}
\label{definition_EFIM_2}
If the FIM of a parameter $\bm{\eta} = [\bm{\eta}_1^{\mathrm{T} }  \; \; \bm{\eta}_2^{\mathrm{T} } ]^{\mathrm{T} }$ is specified by 
\begin{equation}
\label{equ:definition_EFIM_2}
\begin{aligned}
\mathbf{J}_{\bm{\eta}}=\left[\begin{array}{cc}
\mathbf{J}_{\bm{\eta}_1\bm{\eta}_1} & \mathbf{J}_{\bm{\eta}_1\bm{\eta}_2} \\
\mathbf{J}_{\bm{\eta}_1\bm{\eta}_2}^{\mathrm{T}} & \mathbf{J}_{\bm{\eta}_2\bm{\eta}_2}
\end{array}\right],
    \end{aligned}
\end{equation}
where $\bm{\eta} \in \mathbb{R}^{N}, \bm{\eta}_{1} \in \mathbb{R}^{n}, \mathbf{J}_{\bm{\eta}_1\bm{\eta}_1}\in \mathbb{R}^{n \times n}, \mathbf{J}_{\bm{\eta}_1\bm{\eta}_2}\in \mathbb{R}^{n \times(N-n)}$, and $\mathbf{J}_{\bm{\eta}_2\bm{\eta}_2}\in$ $\mathbb{R}^{(N-n) \times(N-n)}$ with $n<N$, 
then the EFIM \cite{emenonye2022ris} of  the parameter of interest ${\bm{\eta}_{1}}$ is given by 
\begin{equation}
\label{equ:definition_EFIM_1}
\begin{aligned}
\mathbf{J}_{\bm{\eta}_{1}}^{\mathrm{e}}&= \mathbf{J}_{{\bm{\eta}_{1}} {\bm{\eta}_{1}}}-
\mathbf{J}_{{\bm{\eta}_{1}} {\bm{\eta}_{2}}} \mathbf{J}_{{\bm{\eta}_{2}}{\bm{\eta}_{2}} }^{-1} \mathbf{J}_{{\bm{\eta}_{1}} {\bm{\eta}_{2}}}^{\mathrm{T}}.
    \end{aligned}
\end{equation}
%Note that the term $\mathbf{J}_{{\bm{\eta}_{1}} {\bm{\eta}_{1}}}^{nu} = \mathbf{J}_{{\bm{\eta}_{1}} {\bm{\eta}_{2}}} \mathbf{J}_{{\bm{\eta}_{2}}{\bm{\eta}_{2}} }^{-1} \mathbf{J}_{{\bm{\eta}_{1}} {\bm{\eta}_{2}}}^{\mathrm{T}}$ describes the loss of information about ${\bm{\eta}_{1}}$  due to uncertainty in the nuisance parameters ${\bm{\eta}_{2}}$.
\end{definition}
 Using Definition \ref{definition_EFIM_2}, the EFIM of the parameter vector $\bm{\eta}$ is computed for different parameters of interest. For example, the EFIM when the parameter of interest is $\bm{\zeta} = [\bm{p}_{U}, \Phi_{U}]^{\mathrm{T}}$ is  $\mathbf{J}_{\bm{\zeta}}^{\mathrm{e}} \in \mathcal{R}^{6 \times 6}$. Here, the nuisance parameter is the complex path gain. %The EFIM when the parameter of interest is the destination's position vector $\bm{p}_{U}$ is  $\mathbf{J}_{\bm{p}_{U}}^{\mathrm{e}} \in \mathcal{R}^{3 \times 3}$. With this parameterization, the nuisance parameter is the source's orientation and the complex path gain.  Finally, the EFIM when the parameter of interest is the source's orientation vector $\bm{\Phi}_{B}$ is  $\mathbf{J}_{\bm{\Phi}_{B}}^{\mathrm{e}} \in \mathcal{R}^{3 \times 3}$. With this parameterization, the nuisance parameter is the destination's position and the complex path gain. Finally, the PEB and OEB are computed for the different parameterizations of the EFIM. The PEB when the parameter of interest is the vector $\bm{\zeta}$ is defined as $\text{PEB}_{\bm{\zeta}} = \operatorname{Tr}([\mathbf{J}_{\bm{\zeta}}^{\mathrm{e}}]_{[1:3,1:3 ]})$ and the OEB in this scenario is defined as $\text{OEB}_{\bm{\zeta}} = \operatorname{Tr}([\mathbf{J}_{\bm{\zeta}}^{\mathrm{e}}]_{[4:6,4:6 ]})$. The PEB when the parameter of interest is the vector $\bm{p}_{U}$ is defined as $\text{PEB}_{\bm{\bm{p}_{U}}} = \operatorname{Tr}([\mathbf{J}_{\bm{\bm{p}_{U}}}^{\mathrm{e}}]_{})$ and the OEB when the parameter of interest is the vector $\bm{\Phi}_{B}$ is defined as $\text{OEB}_{\bm{\bm{\Phi}_{B}}} = \operatorname{Tr}([\mathbf{J}_{\bm{\bm{\Phi}_{B}}}^{\mathrm{e}}]_{})$.
\section{Results}
In this section, we use numerical simulations to find out which combinations of position and orientation parameters can be estimated - a parameter, $\bm{\zeta}$, can be estimated if the corresponding EFIM,  $\mathbf{J}_{\bm{\zeta}}^{\mathrm{e}}$, is positive definite \cite{emenonye2022ris}.   We also provide numerical position error bound (PEB) and orientation error bound (OEB) results for the  
case in which the source orientation and destination position are the unknown parameters. Our simulation framework consists of a source whose centroid is located at $\bm{p}_{B} =  [1.5, 1.0, 4.0]^{\mathrm{T}} $ with the orientation angles $\bm{\Phi}_{B} =  [1.1, 2.2, 0.7]^{\mathrm{T}}$. The position vectors are in meters, and the orientation vectors are in radians. The source has $N_B = 100$ antennas and the following number of transmit beams are considered $N_D \in [16,32,48,64]^{\mathrm{T}}$. For each simulation, $T = 20$ symbols are transmitted, and the beamforming matrix  $\bm{F}_{t} \in \mathcal{C}^{N_B \times N_D}$  changes during each of the $T$ transmit symbols. The rows of this beamforming matrix  are selected from a discrete Fourier transform-based (DFT) codebook. The destination is located at $\bm{p}_{U} =  [2.6, 2.15, 5.1]^{\mathrm{T}} $ with the orientation angles $\bm{\Phi}_{U} =  [0.1, 0.2, 0.1]^{\mathrm{T}}$. The Fraunhofer distance indicates that the destination is experiencing near-field propagation. The incorrect case when the far-field model is applied in this near-field simulation setup is termed ``far-field." The correct case when the near-field model is used is termed ``near-field."

\begin{table*}[h!]
\caption{Location estimation possibilities with near and far field models. N/A indicates the parameter is known. The terms $3$D and $2$D indicate the dimensions in which the parameter can be estimated in the propagation regimes. }
\centering
\begin{tabular}
{|l|l|l|l|l|l|l|l|l|}
\hline & \multicolumn{4}{|c|}{ Near-field } & \multicolumn{4}{c|}{ Far-field  } \\
Unknown Parameters & $\bm{p}_{U}$ & $\bm{\Phi}_{U}$ & $\bm{p}_{B}$ & $\bm{\Phi}_{B}$ & $\bm{p}_{U}$ & $\bm{\Phi}_{U}$ & $\bm{p}_{B}$ & $\bm{\Phi}_{B}$  \\
\hline 
Source position and source orientation  & N/A & N/A & $3$D & $3$D & N/A & N/A & $2$D & $2$D\\
\hline Source position and destination orientation  & N/A & $3$D & $3$D & N/A & N/A & $2$D & $2$D & N/A \\
\hline
 Destination position and source orientation  & $3$D & N/A & N/A & $3$D & $2$D & N/A & N/A & $2$D \\
 \hline 
Source position   & N/A & N/A & $3$D & N/A & N/A & N/A & $2$D & N/A\\
\hline 
Source orientation  & N/A & N/A & N/A & $3$D & N/A & N/A & N/A & $2$D\\
\hline
Destination position   & $3$D & N/A & N/A & N/A & $2$D & N/A & N/A & N/A\\
\hline 
Destination orientation  & N/A & $3$D & N/A & N/A & N/A & $2$D & N/A &  N/A \\
\hline
\end{tabular}
\label{table_1}
\end{table*}
With this simulation setup, we generate Table \ref{table_1}. This table highlights different combinations of the source and destination location that can be estimated. The ``not applicable" term is used to highlight the fact that the parameter is known.  When the term  $3$D is used, it means that the $3$D version of that parameter can be estimated, and if the $3$D version of the parameter can be estimated, all lower dimensions can also be estimated. As evident in Table \ref{table_1}, it is impossible to estimate either the $3$D position coordinates or the $3$D orientation angles with only the signal from the line of sight (LOS) path  when the far-field model is incorrectly applied to the near-field setup. However, if the near-field setup is correctly applied, estimating the $3$D position coordinates or the $3$D orientation angles are feasible with the LOS signal even without a beamforming matrix. While a $2$D estimation of the source or destination's orientation angles is feasible when the far-field model is used and $N_U > 1$, it is important to note that estimating the  source orientation angles is only possible in the far-field with beamforming (see Appendix \ref{appendix_joint_estimation}). This is in contrast with the near-field setup in which the estimation of the source's orientation angles is possible even with no beamforming provided that $N_U > 1$.
\begin{figure}[htb!]
\centering
%\subfloat[]{\includegraphics[height=1.8in, width= 1.5in]{Results/ber_ring_ratio_parallel.eps}
\subfloat[]{\includegraphics[width=\linewidth]{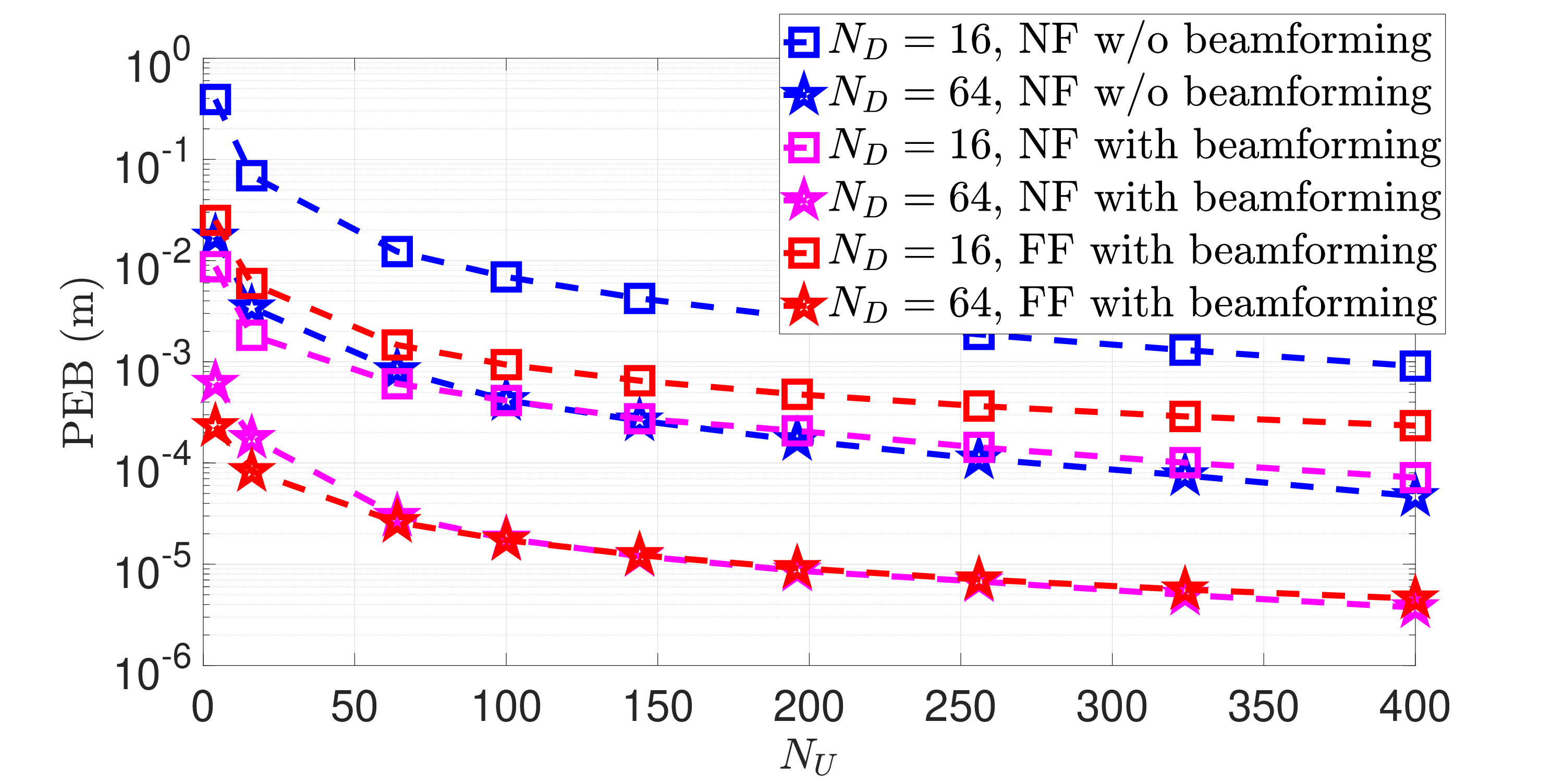}
\label{fig:Figure/PEB_vs_NU}}
\hfil
\subfloat[]{\includegraphics[width=\linewidth]{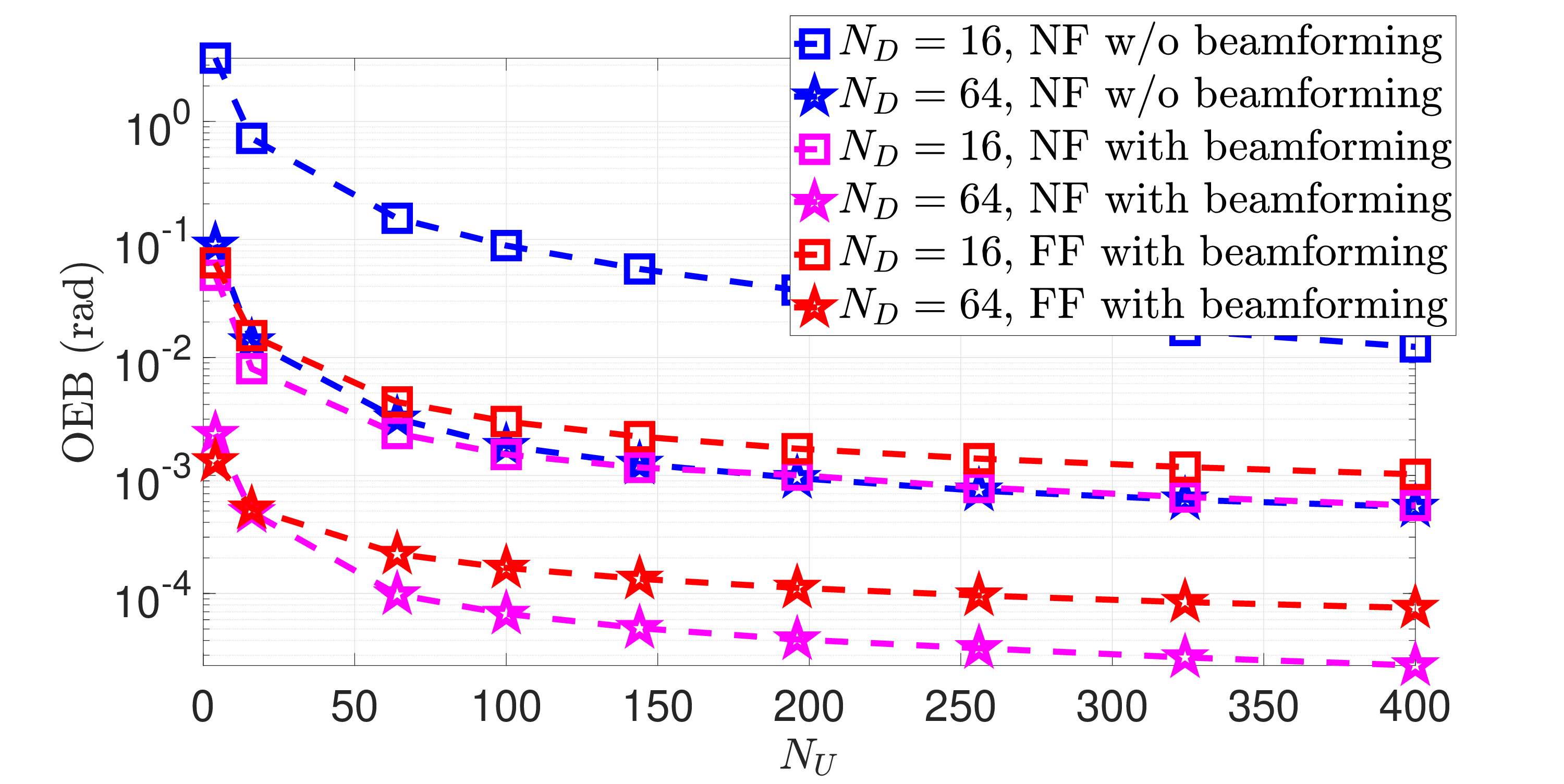}
\label{fig:Figure/OEB_vs_NU}}
\caption{PEB and OEB vs. $N_U$}
\end{figure}
In Figs. \ref{fig:Figure/PEB_vs_NU} and \ref{fig:Figure/OEB_vs_NU}, we present the PEB and OEB as a function of varying numbers of receive antennas. Also, in these figures, the term ``FF" is used to distinguish the incorrect case when the far-field model is applied to the study from the case when the near-field model is correctly applied to the study. As expected, the spherical wavefront in the near-field model results in more accurate localization. From the figures, the spherical wavefront is more advantageous for the estimation of the orientation.
\section{Conclusion}
This paper has examined  the estimation of different combinations of a single-source and single destination's position and orientation. Through a study of the FIM, we have shown that while the source or destination's $3$D orientation can be jointly estimated with the source or destination's $3$D position in the near-field propagation regime, only the source or destination's $2$D orientation can be jointly estimated with the source or destination's $2$D position in the far-field propagation regime. Also, while without beamforming in the near-field, the source's $3$D orientation can be estimated,  the source's $2$D orientation angles can not be estimated without beamforming in the far-field. Finally, a simulation of the PEB and OEB shows that the spherical information present in the near-field is much more useful for estimating orientation information.
\appendix 
\subsection{Analysis of Joint Estimation of $[\bm{\Phi}_{B}, \bm{\beta}]$ under the Far-Field Model}
\label{appendix_joint_estimation}
We start the proof by dropping the subscript $t$ and using  the identity beamforming matrix across the $T$ transmissions. The FIM, $\mathbf{J}_{ \bm{\Phi}_{B}}$, under the parameterization  $\bm{\eta} = [ \bm{\Phi}_{B}, \bm{\beta}
]^{\mathrm{T}}$, is obtained by using the appropriate first derivatives in Definitions (\ref{definition_FIM_1_fundamentals}),  and it has the following structure
$$
\begin{aligned}
\mathbf{J}_{\bm{\eta}} =\left[\begin{array}{ccccc}
\mathbf{J}_{\bm{\Phi}_{B}} & 
 \mathbf{J}_{[\bm{\Phi}_{B}, {\beta}_{\mathrm{R}}]}&  \mathbf{J}_{[\bm{\Phi}_{B}, {\beta}_{\mathrm{I}}]} \\
\mathbf{J}_{ {\bm{\Phi}}_{R},{{\beta}}_R}^{\mathrm{T}} & 
\mathbf{J}_{ {{\beta}}_R} &  0 \\
\mathbf{J}_{ {\bm{\Phi}}_{R},{{\beta}}_I}^{\mathrm{T}} & 
0 &  \mathbf{J}_{ {{\beta}}_I} \\
\end{array}\right],\end{aligned}
$$
and the EFIM  can be written as 
\begin{align}
    \begin{split}
\mathbf{J}_{\bm{\Phi}_{B}}^{\mathrm{e}}  &= \mathbf{J}_{\bm{\Phi}_{B}}- [\mathbf{J}_{ {{\beta}}_R} ]^{-1}\mathbf{J}_{[\bm{\Phi}_{B}, {\beta}_{\mathrm{R}}]}\mathbf{J}_{[\bm{\Phi}_{B}, {\beta}_{\mathrm{R}}]}^{\mathrm{T}} + \mathbf{J}_{[\bm{\Phi}_{B}, {\beta}_{\mathrm{I}}]}\mathbf{J}_{[\bm{\Phi}_{B}, {\beta}_{\mathrm{I}}]}^{\mathrm{T}},
\end{split}
\end{align}
and $
\mathbf{J}_{\bm{\Phi}_{B}}^{\mathrm{e}} = \mathbf{J}_{\bm{\Phi}_{B}}- [\mathbf{J}_{ {{\beta}}_R} ]^{-1}\mathbf{J}_{ {{\beta}}_{R}}\mathbf{J}_{ {\bm{\Phi}}_{R}},$
the second equation results from noticing $\mathbf{J}_{ {{\beta}}_{R}}\mathbf{J}_{ {\bm{\Phi}}_{R}} = \mathbf{J}_{[\bm{\Phi}_{B}, {\beta}_{\mathrm{R}}]}\mathbf{J}_{[\bm{\Phi}_{B}, {\beta}_{\mathrm{R}}]}^{\mathrm{T}} + \mathbf{J}_{[\bm{\Phi}_{B}, {\beta}_{\mathrm{I}}]}\mathbf{J}_{[\bm{\Phi}_{B}, {\beta}_{\mathrm{I}}]}.$ The proof follows as $\mathbf{J}_{\bm{\Phi}_{B}}^{\mathrm{e}} = 0$. Hence, with no beamforming, the source orientation can not be estimated with the far-field propagation model.
{
\bibliographystyle{IEEEtran}
\bibliography{refs}
}
\end{document}